\documentclass[epsfig,12pt,number,sort&compress,onecolumn]{article}

\usepackage[utf8]{inputenc}
\usepackage[english]{babel}
\usepackage[T1]{fontenc}
\usepackage{amsmath}
\usepackage{hyperref}
\usepackage[english]{babel}
\usepackage{amsfonts}
\usepackage{amsthm}
\usepackage{amssymb}
\usepackage{mathtools}
\usepackage{mathrsfs}
\usepackage{pifont}
\usepackage{subcaption}
\usepackage{graphicx}
\usepackage{xspace}
\usepackage[most]{tcolorbox}
\usepackage{color}
\usepackage{braket}
\usepackage{physics}
\usepackage{geometry}
\usepackage{epigraph}
\usepackage{blindtext}
\usepackage[sort,numbers]{natbib}
\usepackage{tikz}
\usepackage{lipsum}
\usepackage{amsfonts}
\usepackage{float}
\usepackage{multicol}
\usepackage{graphicx}
\usepackage{caption}

\makeatletter \@addtoreset{equation}{section}

\def \be{\begin{equation}}
\def \ee{\end{equation}}
\def \bea{\begin{eqnarray}}
\def \eea{\end{eqnarray}}

\newcommand{\nc}{\newcommand}
\nc{\al}{\alpha} \nc{\bib}{\bibitem} \nc{\la}{\lambda}
\nc{\C}{\mbox{\hspace{1.24mm}\rule{0.2mm}{2.5mm}\hspace{-2.7mm} C}}
\nc{\R}{\mbox{\hspace{.04mm}\rule{0.2mm}{2.8mm}\hspace{-1.5mm} R}}

\begin{document}

\title{Quantum Entanglement in the Rabi Model with the Presence of the $A^{2}$ Term}
\author{Z. Boutakka,  Z. Sakhi and M. Bennai$^{1}$\thanks{% 
zakaria.boutakka-etu@etu.univh2c.ma\\
zb.sakhi@gmail.com\\
mohamed.bennai@univh2c.ma} \\
%EndAName
$^{1}${\small Physics and Quantum Technology team, LPMC}\\
{\small \ Ben M'sik Faculty of Sciences, Hassan II University, Casablanca,
Morocco}}
\maketitle

\begin{abstract}
The quantum Rabi model (QRM) is used to describe the light-matter interaction at the quantum level in Cavity Quantum Electrodynamics (Cavity QED). It consists of a two-level system (atom or qubit) coupled to a single-mode quantum field, and by introducing an atom into a cavity alters the electromagnetic mode configuration within it. In the realm of Cavity QED, a notable consequence of this alteration is the emergence of a gauge-dependent diamagnetic term referred to as the $A^{2}$ contribution.
In this study, we comparatively analyze the behaviors of the QRM and the influence of the $A^{2}$ term in the light-matter quantum Hamiltonian by examining the energy spectrum properties in the strong-coupling regime. We then investigate the ground state of the system, measuring its nonclassical properties via the Wigner distribution function for different photon number distribution in Fock space. 
Finally, we calculate the quantum entanglement in the ground state over the Von Neumann entropy.
Our findings reveal that the $A^{2}$ term and the number of cavity Fock states $N$ significantly impact the amount of the quantum entanglement, highlighting their pivotal role.

\end{abstract}

\textbf{Keywords:} cavity $QED$, Quantum Rabi Model, Diamagnetic Term, Wigner Function, Entanglement

\section{Introduction}
Cavity quantum electrodynamics (cQED) is a branch of physics that studies light-matter interaction between restricted optical fields and mechanical degrees of freedom at the entire quantum level. Rabi has presented 80 years ago a model to discuss the effect of a weak, rapidly varying magnetic field on a oriented atom with a nuclear spin \cite{Rabi-1936}. The well-established Rabi model describes the simplest class of light-matter interaction, the dipolar coupling between a two-level quantum system (qubit) and a conventional monochromatic radiation field (one-dimensional harmonic oscillator) \cite{Frisch-1933}. In its quantum version, the radiation is specified by a quantified single-mode field, which gives the Rabi quantum model (QRM) through which we can explains the interaction between a two-level quantum system and a single mode of a bosonic field \cite{Rabi-1936,Xie-2017}.\\ 
It's ironic, how difficult it is to produce a harmonic oscillator state with a precise excitation number $N$, but is not impossible to create Fock states (eigenstates of the harmonic oscillator) that could be achieved by coupling a nonlinear system, such as an atom, to the electromagnetic field. While its excited states are photons and a Fock state corresponds to the creation of n photons with the same energy $\hbar \omega$. \cite{Brune-2008, Hammani}\\
	
Jaynes and Cummings introduced a similar quantum model in 1963 describing a two-level atom interacting with a quantum mode of an optical cavity. Their initial objective was to study the relationship between quantum radiation theory and the corresponding semi-classical theory. Despite its simplicity, Rabi’s quantum model was not considered perfectly resolvable. The model was resolved using a rotating wave approximation (RWA) \cite{Jaynes-1963}, known as the Jaynes–Cummings (JC) model. 
This model accurately describes the dynamics of a wide variety of physical configurations, ranging from quantum optics \cite{Thompson-1992,Marlan-1997,Crespi-2012} and solid state \cite{Toida-2013, Yoshihara-2018}  to condensed matter systems \cite{Holstein-1959} passing through molecular \cite{Kmetic-1990,Albert-2012}. Typically, standard experiments on cavity QED are limited to a much smaller light-matter coupling force than the natural frequencies of undisturbed parts. Thus, they occur in the field of the famous Jaynes-Cummings (JC) model.\\
Nevertheless, the QRM has been applied in cavity QED \cite{Miller-2005}, quantum dots \cite{Hanson-2007}, circuit QED \cite{Wallraff-2004}, and trapped ions \cite{Meekhof-1996}, among other quantum platforms like the development of quantum gates \cite{Schmidt-Kaler-2003,Moya-Cessa-2016,cambridge}.
Furthermore, the advent of quantum information science and the emergence of these novel quantum technologies have led us to investigate the QRM experimentally in depth, including previously inaccessible phenomena such as those occurring in the ultra-strong \cite{Yoshihara-2016} and deep strong \cite{Boité-2016} coupling regimes.\\

An important ingredient is the diamagnetic term (also referred as ‘$A^{2}$ term’) , which is proportional to the square of the vector potential $A^{2}$  and ensures gauge invariance in the non-relativistic minimal coupling Hamiltonian \cite{Woolley-1980}. We note that the $A^{2}$ term is a gauge-dependent object, and specifically appears in the Coulomb gauge description of the single-mode atomic cQED \cite{Moein-2016}.\\
The $A^{2}$ vector potential is a key description of the interaction of atomic matter with bosonic fields in the Coulomb Gauge. This diamagnetic behavior, is caused by the minimal coupling (Gauge-Coulomb invariance), and is derived from the kinetic portion of the Hamiltonian. The interaction of a quantized bosonic mode with non-relativistic charged particles is described by this phrase, which derives from the Coulomb gauge minimal coupling Hamiltonian. It is well known that the inclusion of the D-term in Hamiltonian systems provides a rich discussion centered on the occurrence of the superradiant phase transition.\\
Nataf \cite{Nataf-2010} demonstrated in the context of circuit QED that this term could be avoided by relevantly setting the circuit parameters.  On the other side, in the cavity QED, the inclusion of such a term has been suggested by various authors \cite{Nataf-2010,Rzaewski-1991,Bamba-2014}, and its importance in restricting the effectively possible light-matter coupling \cite{Liberato-2014} and avoiding the superradiant phase transition \cite{Bialynicki-Birula-1979, Chen YH}.\\
The effects associated with this term, are of crucial importance in the research on the ‘Dicke phase transition’, and are still under active investigation and debate \cite{Nataf-2010, Viehmann-2011, Ciuti-2012, Bamba-2014, Liberato-2014}. One of those effects is manifested in the dynamics of the sub-systems, and, most critically, the entanglement phenomena is mysterious \cite{Sakhi-2019}.\\

It is worth mentioning that the diamagnetic term effects can be, in both theoretical predictions and actual experimental studies. Where the diamagnetic term and the concepts developed in the Quantum Rabi model could potentially lead to new discoveries and applications in the constantly growing field of quantum computing \cite{Mirrahimi-2014, Grimm-2020, Bergmann-2016, Xiang-2020},  and information processing \cite{Linshu-2017, V. Albert-2016, Feng-2021}, quantum optics \cite{Crisp-1992} and cavity QED \cite{Ashhab-2010, Leroux-2017}, and quantum simulation and quantum metrology \cite{Jaewoo-2011, Jacob-2021, Mingjie-2021}. All these possible applications stem from the fact that the model helps physicists understand the interaction between a quantum system and an oscillating field, and the effects of the diamagnetic term.

In this paper, we will be using QuTiP \cite{qutip}, that allows us to simulate quantum systems and study  in a comparative way the QRM by neglecting and including the $A^{2}$ term for different cavity Fock number states $N$. This paper is structured as follows:
In Sec. \ref{Sec 2}, we introduce two approaches of QRMs, and we discuss the role of the diamagnetic term in our QRM with the D-Term. In Sec. \ref{Sec 3}, delves into the impact of the diamagnetic term on the energy spectrum and system dynamics. We illustrate this by presenting the transition of the Wigner function distribution from a Schrödinger-cat to a squeezed state in the phase-space, and measuring the quantum entanglement for the ground state across different number of Fock states $N$ included in the cavity. Finally, we conclude with some remarks in Sec. \ref{Sec 4}.
	
\section{Models}
\label{Sec 2}

\subsection{Quantum Rabi Model}
The QRM which describes the important interaction between a two-level system and a quantized electromagnetic field, typically in the form of a cavity field, is expressed by the Hamiltonian :

\begin{equation}
		H_{rabi} = \hbar \omega_{c} a^{\dagger}a + \frac{\hbar \omega_{0}}{2} \sigma_{z} + \hbar g \sigma_{x} (a^{\dagger}+a)
        \label{QRM}
\end{equation}

where $a^{\dagger}$ and $a$ are the field creation and annihilation operators, $\omega_{c}$ is the cavity field frequency, $\sigma_{x}$ and $\sigma_{z}$ are Pauli matrices, $\omega_{0}$ is the transition frequency ($\omega_{0} =  \frac{E_{e} - E_{g}}{ \hbar}$; $E_{e}$ and $E_{g}$ are the energies of the excited and fundamental states, respectively), and $g$ indicates the coupling between the two systems.

\subsection{Quantum Rabi Model with D-Term}

The diamagnetic term can be incorporated into the QRM Hamiltonian to account for the interaction between the charged particle (associated with the two-level system) and the magnetic field. The diamagnetic interaction can be described by a term proportional to the square of the magnetic field, \( B^2 \), and can be written in terms of the vector potential \( A \) as:

\begin{equation}
 H_{\text{dia}} = -\frac{e^2}{2m} A^{2} = D(a + a^{\dagger}) ^{2} 
\end{equation}
Where $e$ is the elementary charge, $m$ is the mass of the charged particle, and the diamagnetic coupling constant $D$ approximately given by $D=\frac{g^{2}}{\omega_{c}}$ \cite{Frisk-2019}.

Incorporating the diamagnetic term into the Rabi model Hamiltonian, the total Hamiltonian becomes:
\begin{equation}
		\begin{aligned}
			H &= H_{Rabi} + H_{dia}\\
			&= \hbar \omega_{c} a^{\dagger}a + \frac{\hbar \omega_{0}}{2} \sigma_{z} + \hbar g \sigma_{x} (a^{\dagger}+a) + D (a^{\dagger}+a)^{2}
		\end{aligned}
	\label{H dia}
\end{equation}

Therefore, the diamagnetic term $H_{\text{dia}}$ would add an additional term to the Hamiltonian of the system, representing the interaction between the two-level system and the magnetic field.

\section{Results \& Discussion}
\label{Sec 3}

The diamagnetic term affects the system's energy levels and can induce alterations in the energy spectrum and dynamics of the Quantum Rabi Model, especially in situations where the magnetic field holds prominence.\\
Hence, in the subsequent, we investigate the energy spectrum and dynamics of the two approaches model as a function of the coupling strength between the qubit and the cavity field, under resonance conditions $\omega_{c}=\omega_{0}$ and for different values of $N$.

\subsection{Energy Spectrum}
Let's consider that the QRMA, i.e., the quantum Rabi model (QRM) with the $A^2$ term.\\
In Figures 1 and 2, we display a comparison of the eigenvalues found in the frameworks of QRM and QRMA solutions as a function of a coupling constant $g$. 

When the parameter $D$ is set to zero, the solution of the QRM in the strong coupling regime differs dramatically from the QRMA, as expected and well known.\\

  \begin{figure}[ht]
		\begin{subfigure}{0.5\textwidth}
		\includegraphics[width=0.8\linewidth]{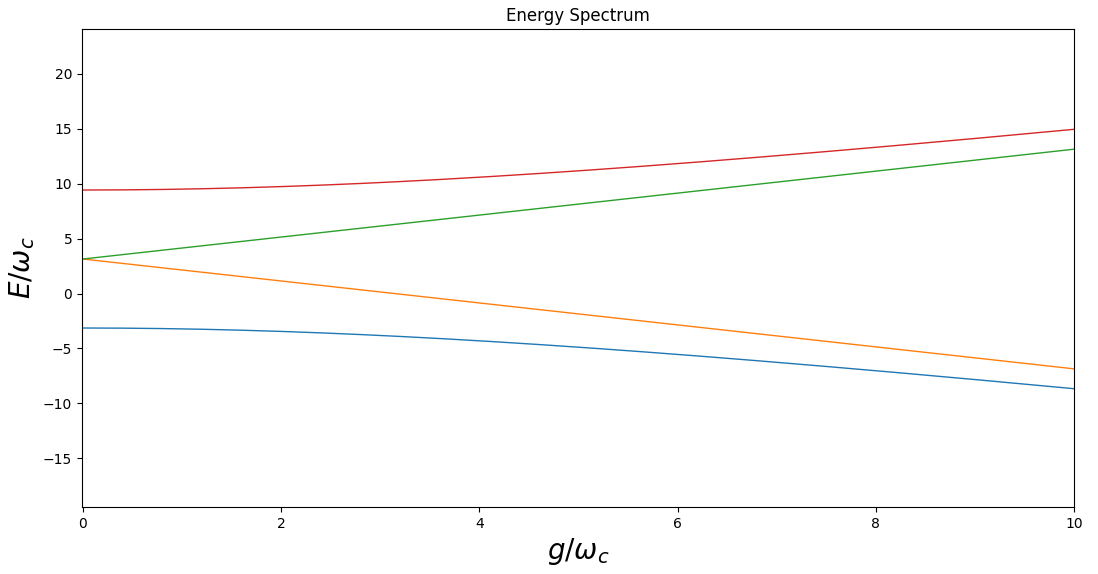}  
	\caption{}		
        \label{fig1a}
        \end{subfigure}
        \hfill
		\begin{subfigure}{0.5\textwidth}
		\includegraphics[width=0.8\linewidth]{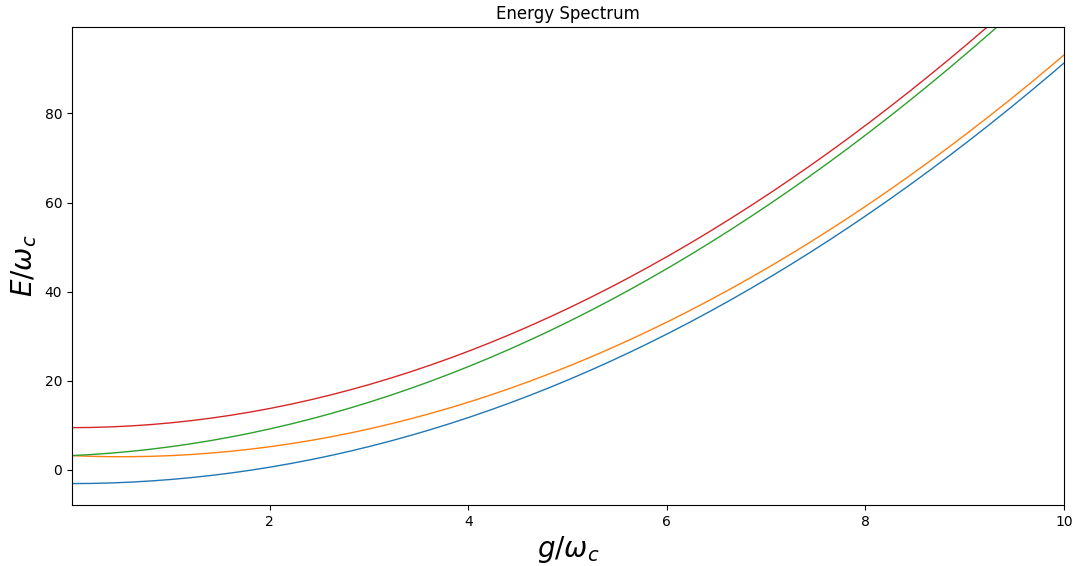}
        \caption{}
		\label{fig1b}
        \end{subfigure}
  
    \caption{Energy eigenvalue spectra of the two approaches of the QRM plotted as a function of the coupling strength $\frac{g}{\omega_{c}}$ for $\frac{\omega_{0}}{\omega_{c}} = 1$ and number of cavity Fock states $N=2$.}
    \label{fig1}
    \end{figure}
    
For $n=2$, the figure \ref{fig1} shows that the variation of the eigenvalues of the quantum Rabi model in relation to $\frac{g}{\omega_{c}}$, which split into two regions and that indicates that we deal with a two level system “qubit”.\\
We note that the quantum Rabi model’s own eigenvalues are not degenerated, which leads to say that the counter resonant terms (CRT) remove degeneration.\\

 \begin{figure}[ht]
		\begin{subfigure}{0.5\textwidth}
		\includegraphics[width=0.8\linewidth]{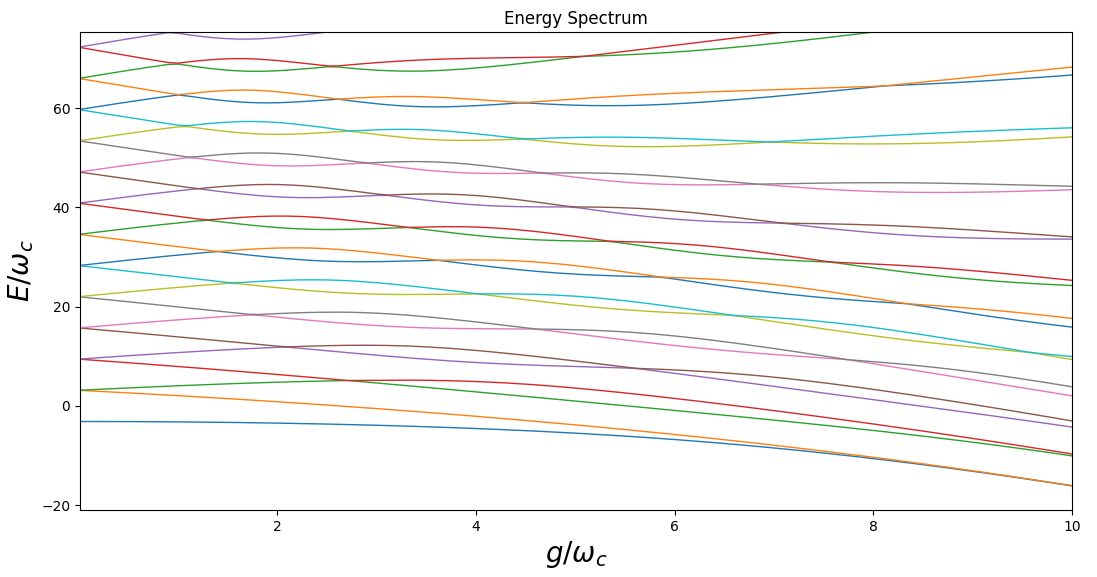}  
	\caption{}		
        \label{fig2a}
        \end{subfigure}
        \hfill
		\begin{subfigure}{0.5\textwidth}
		\includegraphics[width=0.8\linewidth]{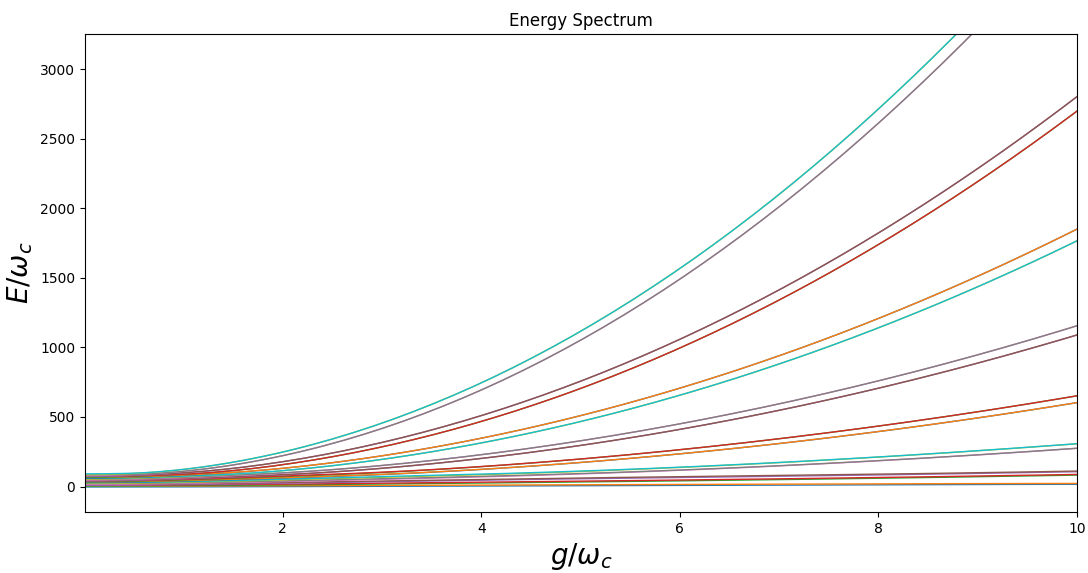}
        \caption{}
		\label{fig2b}
        \end{subfigure}
  
    \caption{The energy spectrum of the two approaches models versus the coupling strength $\frac{g}{\omega_{c}}$ for $\frac{\omega_{0}}{\omega_{c}} = 1$ and number of cavity Fock states $N=15$.}
    \label{fig2}
    \end{figure}

In figure \ref{fig2a}, when $N > 2$, we observe points where the energy levels of the system come close to each other without actually crossing. These are called avoided level crossings.
Simply, when the curves of energy repel each other, causing them to avoid crossing and maintain a certain minimum energy separation. \\
This phenomenon arises due to the non-commutativity of the operators representing different physical quantities in quantum mechanics, it signifies a coupling between the qubit and the cavity field, where their energies become entangled and influence each other. Specifically, when the energies of the two different eigenstates of the system approach each other, instead of crossing. 

\begin{figure}[!h]
		\begin{subfigure}{0.5\textwidth}
		\includegraphics[width=0.8\linewidth]{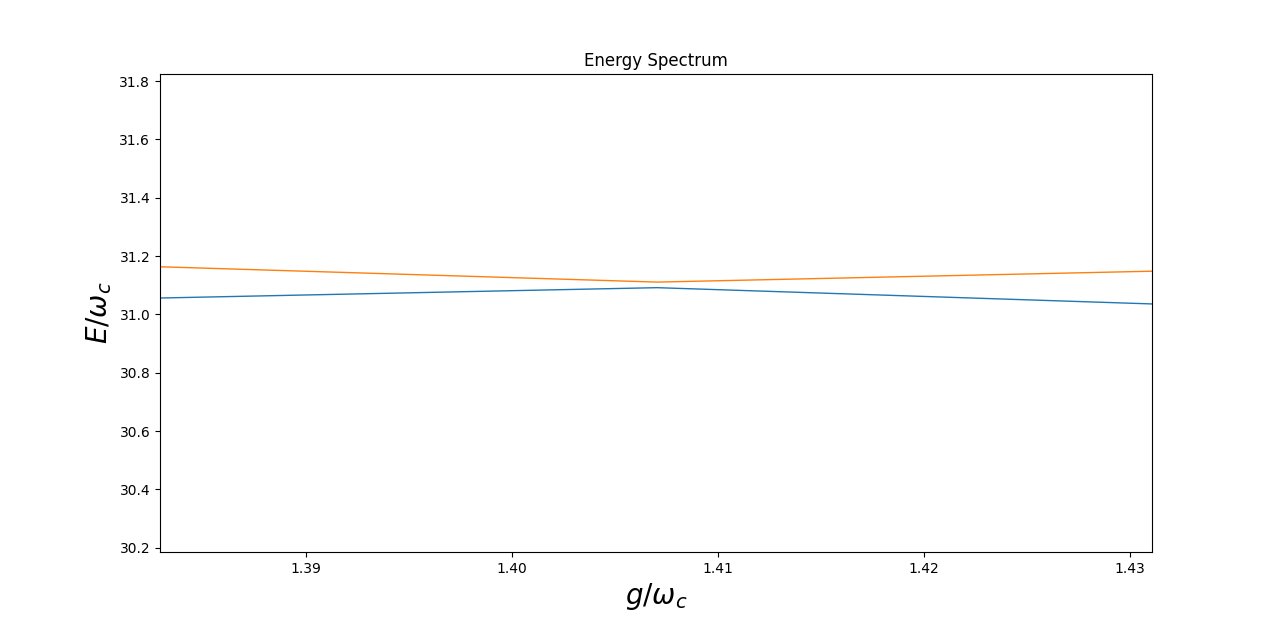}  
	\caption{}		
        \label{}
        \end{subfigure}
        \hfill
		\begin{subfigure}{0.5\textwidth}
		\includegraphics[width=0.8\linewidth]{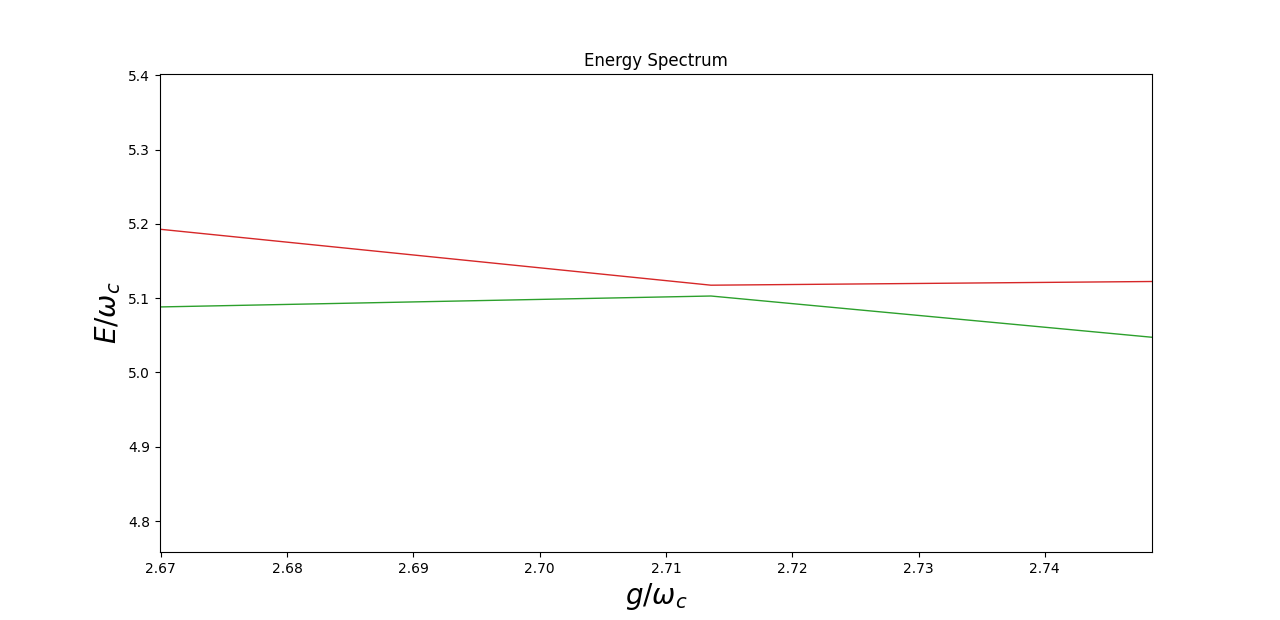}
        \caption{}
		\label{}
        \end{subfigure}
        \begin{subfigure}{0.5\textwidth}
		\includegraphics[width=0.8\linewidth]{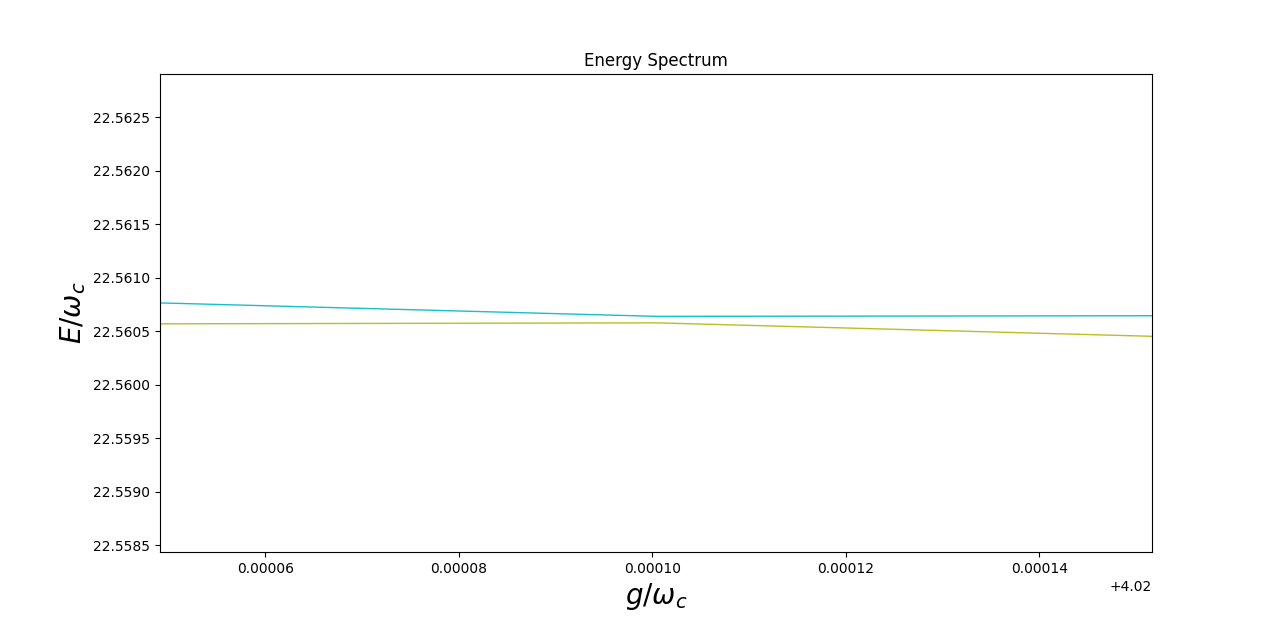}  
	\caption{}		
        \label{}
        \end{subfigure}
        \hfill
		\begin{subfigure}{0.5\textwidth}
		\includegraphics[width=0.8\linewidth]{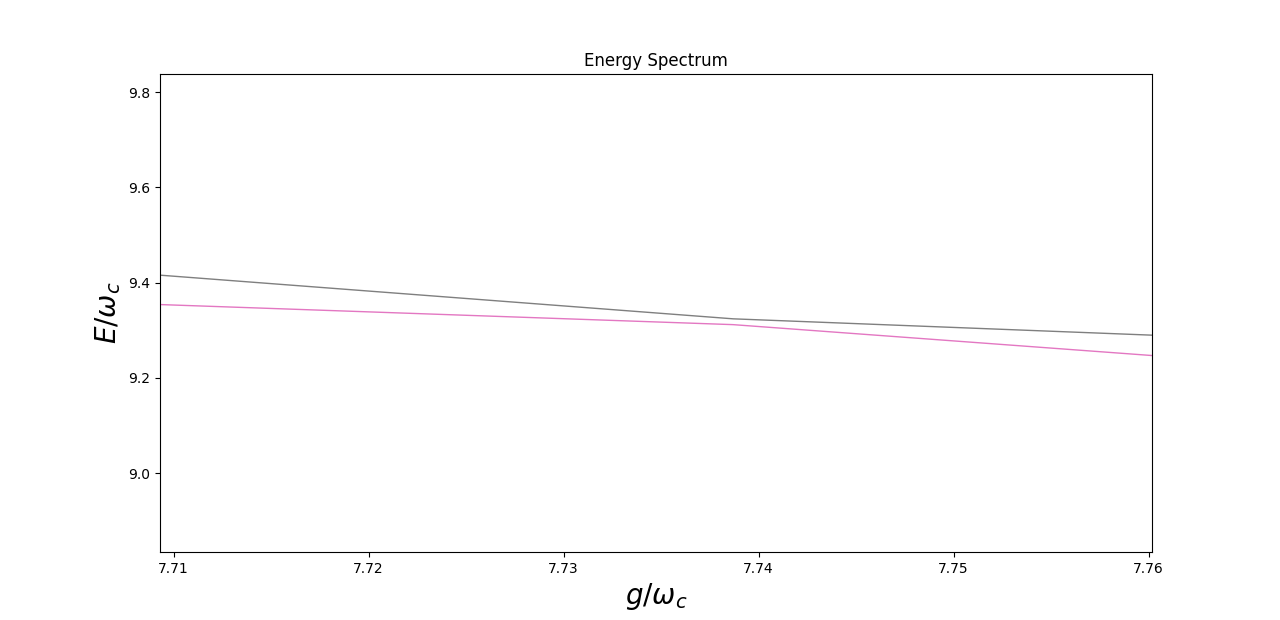}
        \caption{}
		\label{}
        \end{subfigure}
		\caption{Zoom in of the spectrum of the QRM in figure \ref{fig2a} , showing the avoided level crossing at different values of the coupling strength $\frac{g}{\omega_{c}}$.}
		\label{fig3}
\end{figure}

The avoided level crossing phenomenon is shown clearly in the fig \ref{fig3}, where we can take any random point from the energy spectrum exhibited in the fig \ref{fig2a}, we may observe that at this point the energy do not cross due to a competition between resonant and counter resonant terms.

While in the figures \ref{fig1b} and \ref{fig2b}, it is imperative to underscore the appearance of the diamagnetic term related to the condition $D \neq 0$, as it exerts a notable influence on the spectrum of the system. Through our observation, we discern that varying the cavity number $N$ results in alterations to the energy spectrum due to the presence of the $A^{2}$ term in the Coulomb gauge. This phenomenon, characterized by the modified spectral properties, is denoted as the diamagnetic shift energy.
Moreover, due to the $A^{2}$ term, this requirement makes it significantly more difficult for a QRMA with purely electromagnetic interaction to enter the strong coupling regime, then the degeneracy of the QRMA is observed to be important for considerably greater values of the coupling constant $g$.

\subsection{Wigner Function}

The Wigner function was introduced by Wigner \cite{Wigner-1932} as a useful tool to express quantum mechanics in a phase space formalism.  It is very suitable for analyzing the transition from classical to quantum dynamics. It provides a specific description of some measurements performed on quantum states, but this is done at the expense of being potentially negative. Therefore, the Wigner function cannot be interpreted as a regular probability distribution for a random variable, but it is generally interpreted as a quasi-probability distribution.\\
	
For any $\ket{\psi}$ state in one-dimensional Hilbert space, the state can be represented as a complex wave function:
$\ket{\psi}= \int \psi(q) dq \ket{q}$ where $\braket{q}{\psi} =  \psi(q)$.  Similarly, a mixed general state is expressed as a matrix of density $\rho = \int \rho(q_{1}, q_{2}) \braket{q_{1}q_{2}}$. \\
	
The Wigner $W(q, p)$ function provides an equivalent representation of any quantum state in the quadrature phase space $(q, p)$. In the general case, which includes mixed states, the Wigner function is defined as
	\begin{equation}
		W(q,p) = \frac{1}{2\pi \hbar} \int_{-\infty}^{\infty}  \mel{q+ \frac{\mu}{2}}{\rho}{q- \frac{\mu}{2}} e^{\frac{- i p \mu}{\hbar}} d\mu 
	\end{equation}

With variable change $y = - \frac{\mu}{2}$, and taking $\hbar=1$ we can find the following equivalent definition	
	\begin{equation}
		W(q,p) = \frac{1}{\pi} \int_{-\infty}^{\infty}  \mel{q- y}{\rho}{q+y} e^{2ipy} dy 
		\label{wigner fct}
	\end{equation}
	
where, $q$ \& $p$ are the position and momentum parameters.\\
	
The expression (\ref{wigner fct}) is not the only formula; there are other formulations for the Wigner function. Here, we will define the interesting one
	
	\begin{equation}
		W(\gamma) = \frac{1}{\pi^{2}} \int  C_{W}(\lambda) e^{\lambda^{*}\gamma - \lambda\gamma^{*}} d^{2}\lambda 
	\end{equation}
	
In which $C_{W}(\lambda) =  \Tr(\rho D(\lambda))$ is a characteristic function of Wigner and $D(\lambda) = e^{(\lambda a^{\dagger} - \lambda^{*} a)} $ is the displacement operator. $\rho$ is the reduced density matrix obtained by partially trace on the second mode.\\

For numerous reasons, we will employ the Wigner function in the following: First and foremost, it is appropriate for studying the transition from classical to quantum dynamics \cite{Siyouri}. Second, it provides a rich framework for representing quantum dynamics purely in the language of phase space variables. Finally, it is sensitive to phase space interference and so provides a clear forecast of the possibility of nonclassical consequences of the quantum mechanical system (accurate benchmark of the quantum effects.)\cite{Polkovnikov-2010}. \\
	
\begin{figure}
		\begin{subfigure}{1\textwidth}
			\centering
			% include first image
			\includegraphics[width=1\linewidth]{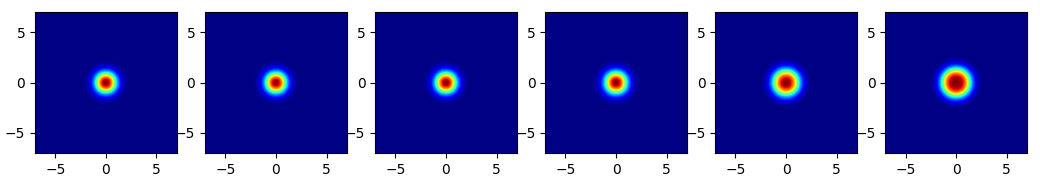}  
                \caption{}
			\label{fig4a}
		\end{subfigure}
  
		\begin{subfigure}{1\textwidth}
			\centering
			% include second image
			\includegraphics[width=1\linewidth]{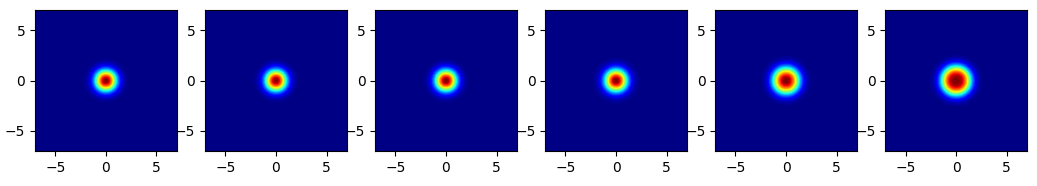}
                \caption{}
			\label{fig4b}
		\end{subfigure}
		
\caption{Phase Momentum presentation of the ground state wave function for the QRM without and with the $A^{2}$ term for different values of the coupling strength ($g = 0 , 0.5 , 1 , 3 , 7 , 10$ and $N = 2$)}
\label{fig4}
\end{figure}
 
\begin{figure}
		\begin{subfigure}{1\textwidth}
			\centering
			% include first image
			\includegraphics[width=1\linewidth]{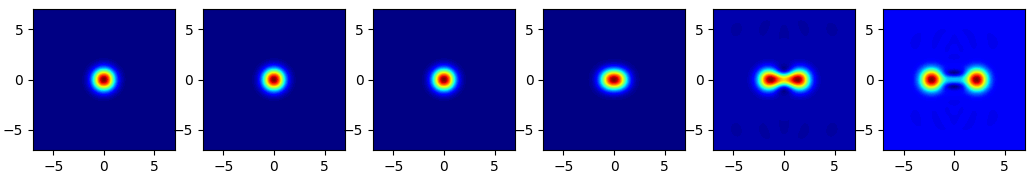}
                \caption{}
			\label{fig5a}

		\end{subfigure}
  
		\begin{subfigure}{1\textwidth}
			\centering
			% include second image
			\includegraphics[width=1\linewidth]{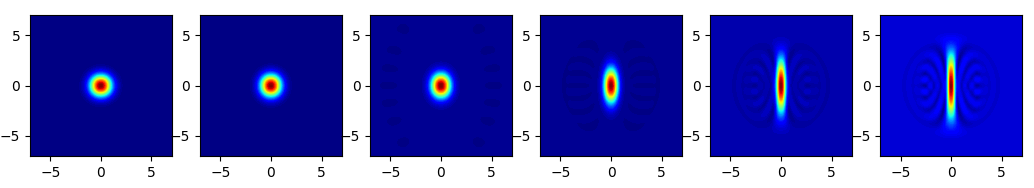}  
			\caption{}
                \label{fig5b}
                
		\end{subfigure}
		
\caption{Phase Momentum presentation of the ground state wave function for the QRM without \ref{fig5a} and  with \ref{fig5b} the $A^{2}$ term for different values of the coupling strength ($g =0 , 0.5 , 1 , 3 , 7 , 10$ and $N = 15$)}
\label{fig5}
\end{figure}
	
We calculate the Wigner function of the ground state based on the Hamiltonian of the both models, as plotted in Fig \ref{fig4} , \ref{fig5}. The quadratures are defined as $x = \frac{(a + a^{\dagger} )}{\sqrt{2}}$ and $y = \frac{(a - a^{\dagger} )}{\sqrt{2i}}$.\\

The Figures \ref{fig4}, \ref{fig5} represent a 2D numerical plot of the Wigner distribution function of the ground state as a function of the \textbf{`position q'} and the \textbf{`momentum p'}.\\

For coupling strength $g$, the ground state of the system is a vacuum state with a Gaussian distribution shown in Fig \ref{fig4}. For $g=0$ and $g\neq0$, we observe that the Wigner distribution function in position and space is concentrated around $q=p=0$. While the $\Delta q$ and $\Delta p$ of the function increase with the augmentation of $g$ along the two axis $(q,p)$.

In Fig \ref{fig5}, we present the Wigner function of the ground states in the cases of ignoring and including the $A^{2}$ term respectively, for a high cavity number ($N=15$). 
First of all, it clearly presents that in the case of ignoring the $A^{2}$ term (QRM), the entanglement and quantum superposition phenomena can be realized when the ground state wave function diverges and splits into two different regions of phase-space, it’s actually a Schrodinger cat state, as shown in Fig \ref{fig5a}. 
By including the $A^{2}$ term in the Wigner distribution function of the ground state is being pressed along the $q$ axis, which made it, stretch in the $p$ direction up to its maximum. Then we obtain a squeezed state as a result of the addition of the $A^2$ term (see Fig \ref{fig5b}).

\begin{figure}[!h]
		\begin{subfigure}{0.5\textwidth}
		\includegraphics[width=0.8\linewidth]{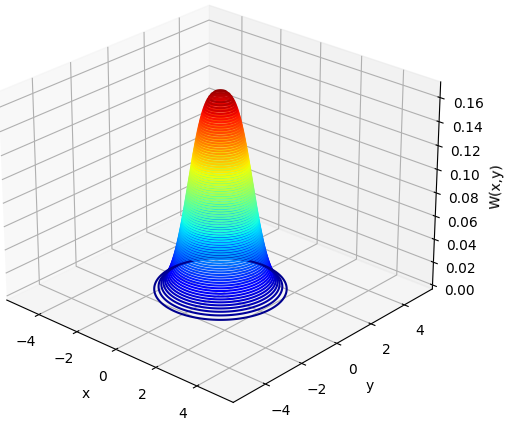}  
		\caption{}
        \label{fig6a}
		\end{subfigure}
        \hfill
		\begin{subfigure}{0.5\textwidth}
        \includegraphics[width=0.8\linewidth]{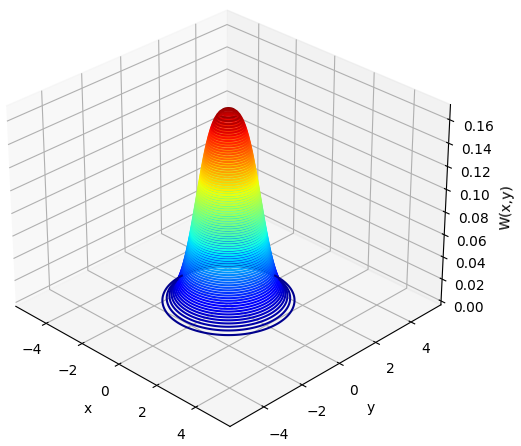}  
		\caption{}
        \label{fig6b}
		\end{subfigure}
  
\caption{3D representation of Wigner function of the ground state wave function for the two models when $g=10 $ and $N = 2$}
\label{fig6}
\end{figure}

\begin{figure}[ht]
		\begin{subfigure}{0.5\textwidth}
		\includegraphics[width=0.8\linewidth]{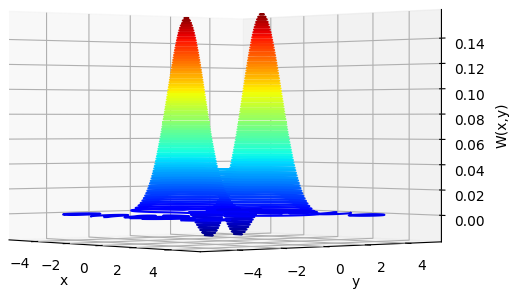}  
	\caption{}		
        \label{fig7a}
        \end{subfigure}
        \hfill
		\begin{subfigure}{0.5\textwidth}
		\includegraphics[width=0.8\linewidth]{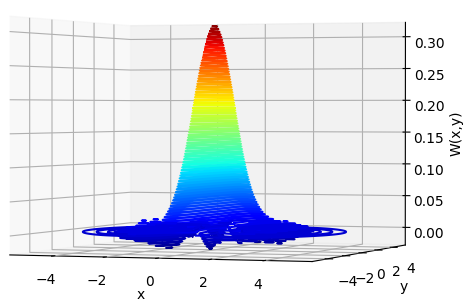}
        \caption{}
		\label{fig7b}
        \end{subfigure}
  
\caption{3D representation of Wigner function of the ground state wave function for the two models when $g = 10 $ and $N = 15$}
\label{fig7}
\end{figure}

The Figures \ref{fig6} and \ref{fig7}  show a 3D visualization of the previous results in the all cases.
    
We visualize that for coupling strength $\frac{g}{\omega_{c}}$, the ground state of the approaches models (including \& neglecting the $A^{2}$ term, respectively) is a vacuum state with a Gaussian distribution shown in the Figures \ref{fig6a},  \ref{fig6b} when $\frac{g}{\omega_{c}}$ takes value $10$ and $N$ limited in $2$.\\

While, when the $N$ is higher than $2$ (in our example $n = 15$), we observe that on the QRM an appearance of some peaks in the center, oscillating between positive and negative values, represent the interference fringes that reveals the coherent superposition of the two components, this state is called “ Schrödinger cat state” shown in Fig \ref{fig7a}. On the other hand, when we include the $A^{2}$ term, the state become a squeezed vacuum state (see Figure \ref{fig7b}).
    
\subsection{Entanglement}

With the aim of differentiate clearly between Schrodinger cat states in the cavity and the system entangled states, we investigate the entanglement properties in the ground state.

Entanglement is a fascinating feature of the quantum mechanics \cite{EPR-1935} and a cornerstone in various fields, such as quantum computation, quantum communication, quantum cryptography, quantum metrology, and quantum simulation. In effect, many concepts and formalisms are existing in purpose to quantify this one \cite{Verdal-1997, Wootters-1998, Vidal-2002}. 

In order to study quantum entanglement dynamics process between the qubit and the cavity field in the two approaches models quantitatively, we numerically calculate the von Neumann entropy $S = -\Tr(\rho\log_{2} \rho)$ of the reduced density matrix $\rho$ of the system

\begin{figure}[!h]
		\begin{subfigure}{0.5\textwidth}
		\includegraphics[width=0.8\linewidth]{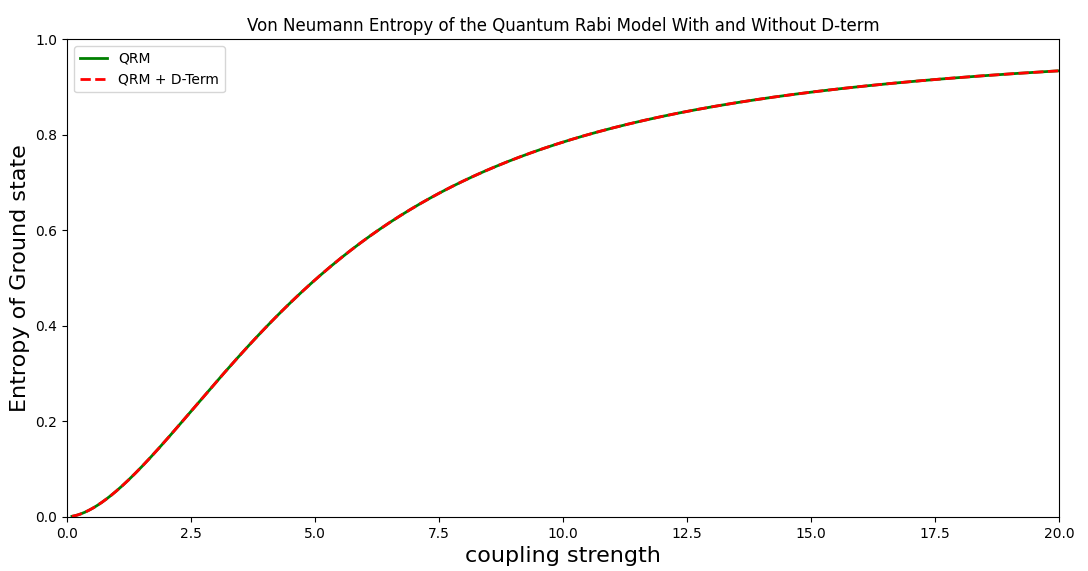}  
		\caption{}
        \label{figa}
		\end{subfigure}
        \hfill
		\begin{subfigure}{0.5\textwidth}
        \includegraphics[width=0.8\linewidth]{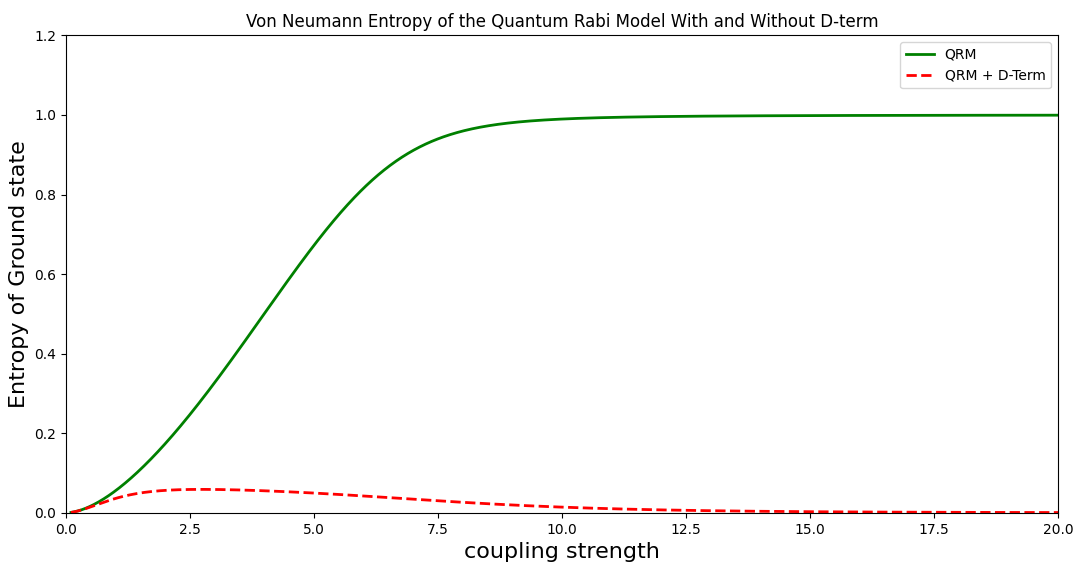}  
		\caption{}
        \label{figb}
		\end{subfigure}
  
\caption{The Von Neumann entropy $S$ which quantifies
the system entanglement in the ground state as a function
of the coupling strength for different number of cavity Fock states : \textbf{(a)} $N = 2$ ; \textbf{(b)} $N = 15$. The QRM (green, solid line) , the QRM in the presence of the diamagnetic term (red, dashed line). }
\label{fig8}
\end{figure}

The numerical results of the Von Neumann entropy (Figure \ref{fig8}) show that for the cavity number states $N$ is equal to $2$, the system is correlated without being maximally entangled for the both approaches models (seen Figure \ref{figa}).

Hence, for large value of the cavity number states $N$ (Figure \ref{figb}), the Schrodinger cat state in the QRM becomes maximally entangled $(D = 0)$ according to the results found in figures above in the phase-space representation, while the entanglement undergoes has a drop or degradation when the diamagnetic term is included $(D \neq 0)$, and it tends towards $0$ wherever the diamagnetic term takes high values.

In that case, the results in Figure \ref{fig8} indicate that the cavity number $N$ and the diamagnetic term $D$ play an effective role when it comes to increasing or limiting  the emergence of entanglement in the ground state of the model.

\section{Conclusion}
\label{Sec 4}

In this work, we have compared the simulations of two different approaches model, by analyzing the energy spectrum properties of the qubit-cavity field under the strong coupling, considering the inclusion and exclusion of diamagnetic effect.

Further, We have presented the numerical simulations of the Wigner function distribution and Von Neumann entropy to measure nonclassicality and entanglement, respectively, which are important for the modern quantum technologies such as quantum computation \cite{Grimm-2020, Bergmann-2016, Xiang-2020}, quantum information processing \cite{Linshu-2017, V. Albert-2016, Feng-2021}, and quantum metrology \cite{Jaewoo-2011, Jacob-2021, Mingjie-2021}.

Moreover, by augmenting the cavity number states $N$ in Fock space for the ground state of the QRM with and without the diamagnetic term, we clearly indicate that the system evaluate through the vacuum states, the squeezed vacuum states, and the Schrödinger-cat states without squeezing as the coupling strength increases \cite{Ashhab-2010, Liu-2024}.

We have also analyzed the effect of the diamagnetic term on the system. We have demonstrated that, the realizations of ground-state quantum entanglement and superposition in normal cavity QED systems would be limited by the existence of the $A^{2}$ term whenever the cavity number states $N$ is greater than $2$.

\begin{comment}
\bibliography{biblio.bib}	

\newpage


\begin{thebibliography}{100}

\bibitem{Rabi-1936} I. I. Rabi Phys. Rev. 49, 324 (1936).

\bibitem{Frisch-1933} R. Frisch, E. Segrè, Über die Einstellung der Richtungsquantelung. II. Z. Physik 80, 610–616 (1933).

\bibitem{Xie-2017} Qiongtao Xie; Honghua Zhong; Murray T Batchelor; Lee Chaohong, arXiv:1609.00434 [quant-ph] submitted (Sep 2016)

\bibitem{Brune-2008} M. Brune, J. Bernu, C. Guerlin, S. Deléglise, C. Sayrin, S. Gleyzes, S. Kuhr, I. Dotsenko, J. M. Raimond, and S. Haroche, Phys. Rev. Lett. 101, 240402, 2008.

\bibitem{Hammani} Hammani M., Sakhi Z., \& Bennai, M
On the two-photon quantum Rabi model at the critical coupling strength.  
Optical and Quantum Electronics, 56(1), 1-14. (2024)

\bibitem{Jaynes-1963}  E. Jaynes and F. Cummings, Proc. IEEE 51, 89 (1963)

\bibitem{Thompson-1992} R. J. Thompson, G. Rempe, and H. J. Kimble,
Phys. Rev. Lett. 68, 1132, 1992.

\bibitem{Marlan-1997} Marlan O. Scully and M. Suhail Zubairy,
\emph{Quantum Optics}
\textrm{(Cambridge University Press, 1997)}

\bibitem{Crespi-2012} A. Crespi, S. Longhi, and R. Osellame,
Phys. Rev. Lett. 108, 163601, 2012.

\bibitem{Toida-2013} H. Toida, T. Nakajima, and S. Komiyama,
Phys. Rev. Lett. 110, 066802, 2013.

\bibitem{Yoshihara-2018} F. Yoshihara, T. Fuse, Z. Ao, S. Ashhab, K. Kakuyanagi, S. Saito, T. Aoki, K. Koshino, and K.
Semba,
Phys. Rev. Lett. 120, 183601, 2018.

\bibitem{Holstein-1959} T. Holstein
Ann. of Phys. (N.Y.)
(1959)

\bibitem{Kmetic-1990} Mary Ann Kmetic and William J. Meath,
Phys. Rev. A 41, 1556

\bibitem{Albert-2012} Victor V. Albert,
Phys. Rev. Lett. 108, 180401, 2012.

\bibitem{Miller-2005} R. Miller et al 2005 J. Phys. B: At. Mol. Opt. Phys. 38 S551

\bibitem{Hanson-2007} R. Hanson, L. P. Kouwenhoven, J. R. Petta, S. Tarucha, and L. M. K. Vandersypen,
Rev. Mod. Phys. 79, 1217, 2007.

\bibitem{Wallraff-2004} Wallraff, A., Schuster, D., Blais, A. et al. Strong coupling of a single photon to a superconducting qubit using circuit quantum electrodynamics. Nature 431, 162–167 (2004).

\bibitem{Meekhof-1996} D. M. Meekhof, C. Monroe, B. E. King, W. M. Itano, and D. J. Wineland,
Phys. Rev. Lett. 76, 1796 (1996).


\bibitem{Schmidt-Kaler-2003} F. Schmidt-Kaler, H. Häffner, M. Riebe, et al. Realization of the Cirac–Zoller controlled-NOT quantum gate. Nature 422, 408–411 (2003).

\bibitem{Moya-Cessa-2016} Héctor M. Moya-Cessa, Fast Quantum Rabi Model with Trapped Ions. Sci Rep 6, 38961 (2016).

\bibitem{cambridge}  M. A. Nielsen and I. L. Chuang, Quantum Computation and Quantum Information
(Cambridge University Press, 2000).

\bibitem{Yoshihara-2016} Fumiki Yoshihara, Tomoko Fuse, Sahel Ashhab et al. Superconducting qubit–oscillator circuit beyond the ultrastrong-coupling regime. Nature Phys 13, 44–47 (2017).

\bibitem{Boité-2016} Alexandre Le Boité, Myung-Joong Hwang, Hyunchul Nha, and Martin B. Plenio,
Phys. Rev. A 94, 033827

\bibitem{Woolley-1980} R. G. Woolley, 1980 J. Phys. A: Math. Gen. 13 2795

\bibitem{Moein-2016} Moein Malekakhlagh and Hakan E. Türeci
Phys. Rev. A 93, 012120 (2016)

\bibitem{Nataf-2010} Pierre Nataf , Cristiano Ciuti. No-go theorem for superradiant quantum phase transitions in cavity QED and counter-example in circuit QED. Nat Commun 1, 72 (2010).

\bibitem{Rzaewski-1991} K. Rzaewski ; K. Wodkiewicz,
Phys Rev A. 1991 Jan 1;43(1):593-594.

\bibitem{Bamba-2014} Motoaki Bamba and Tetsuo Ogawa,
Phys. Rev. A 90, 063825, 2014.

\bibitem{Liberato-2014} Simone De Liberato,
Phys. Rev. Lett. 112, 016401, 2014.

\bibitem{Bialynicki-Birula-1979} Iwo Bialynicki-Birula and Kazimierz Rzażewski,
Phys. Rev. A 19, 301, 1979.

\bibitem{Chen YH} Chen, YH., Qiu, Y., Miranowicz, A. et al. Sudden change of the photon output field marks phase transitions in the quantum Rabi model. Commun Phys 7, 5 (2024).

\bibitem{Viehmann-2011} Oliver Viehmann, Jan von Delft, and Florian Marquardt,
Phys. Rev. Lett. 107, 113602, 2011.

\bibitem{Ciuti-2012} Cristiano Ciuti and Pierre Nataf,
Phys. Rev. Lett. 109, 179301, 2012.

\bibitem{Sakhi-2019} Z. Sakhi, A. Chentouf, M. Bennai, (D. Mogilevtsev) .
\emph{“Effect of the diamagnetic term in the ultra-strong coupling regime,”}
\textrm{EPJ Web of Conferences, vol. 198, p. 00011, 2019.}

\bibitem{Grimm-2020} Grimm, A., Frattini, N.E., Puri, S. et al. Stabilization and operation of a Kerr-cat qubit. Nature 584, 205–209 (2020).

\bibitem{Bergmann-2016} Marcel Bergmann and Peter van Loock
Phys. Rev. A 94, 042332 (2016).

\bibitem{Xiang-2020} Xiang-You Chen, Yu-Yu Zhang, Libin Fu, and Hang Zheng
Phys. Rev. A 101, 033827 (2020)

\bibitem{Mirrahimi-2014} Mazyar Mirrahimi et al 2014 New J. Phys. 16 045014

\bibitem{Linshu-2017} Linshu Li, Chang-Ling Zou, Victor V. Albert, Sreraman Muralidharan, S.M. Girvin, and Liang Jiang
Phys. Rev. Lett. 119, 030502 (2017).

\bibitem{V. Albert-2016} Victor V. Albert, Chi Shu, Stefan Krastanov, Chao Shen, Ren-Bao Liu, Zhen-Biao Yang, Robert J. Schoelkopf, Mazyar Mirrahimi, Michel H. Devoret, and Liang Jiang
Phys. Rev. Lett. 116, 140502 (2016).

\bibitem{Feng-2021} Feng-Xiao Sun, Sha-Sha Zheng, Yang Xiao, Qihuang Gong, Qiongyi He, and Ke Xia
Phys. Rev. Lett. 127, 087203 (2021).

\bibitem{Jaewoo-2011} Jaewoo Joo, William J. Munro, and Timothy P. Spiller
Phys. Rev. Lett. 107, 083601 (2011).

\bibitem{Jacob-2021} Jacob Hastrup, Kimin Park, Radim Filip, and Ulrik Lund Andersen
Phys. Rev. Lett. 126, 153602 (2021).

\bibitem{Mingjie-2021} Mingjie Xin, Wui Seng Leong, Zilong Chen, Yu Wang, and Shau-Yu Lan
Phys. Rev. Lett. 127, 183602 (2021)

\bibitem{Crisp-1992} Michael D. Crisp
Phys. Rev. A 46, 4138 (1992)

\bibitem{Ashhab-2010} S. Ashhab and Franco Nori
Phys. Rev. A 81, 042311 (2010)

\bibitem{Leroux-2017} C. Leroux, L. C. G. Govia, and A. A. Clerk
Phys. Rev. A 96, 043834 (2017)

\bibitem{qutip} J. Johansson, P. Nation, F. Nori,
\emph{“QuTiP: An open-source Python framework for the dynamics of open quantum systems,”}
\textrm{Computer Physics Communications, 183(8), 1760-1772 (2012).}

\bibitem{Frisk-2019} Frisk Kockum, A., Miranowicz, A., De Liberato, S. et al. Ultrastrong coupling between light and matter. Nat Rev Phys 1, 19–40 (2019)

\bibitem{Wigner-1932} E. Wigner,
Phys. Rev. 40, 749 (1932).

\bibitem{Siyouri} Fatimazahra SIYOURI, (2017)
Contributions to the study of quantum correlations. Doctoral thesis (PhD), Mohammed V University of Rabat, 
Retrieved from Presses académiques francophones.

\bibitem{Polkovnikov-2010} Anatoli Polkovnikov,
Ann. of Phys. 325, 1790–1852,
(2010)

\bibitem{EPR-1935} A. Einstein, B. Podolsky, and N. Rosen
Phys. Rev. 47, 777 (1935)


\bibitem{Verdal-1997} V. Vedral, M. B. Plenio, K. Jacobs, and P. L. Knight
Phys. Rev. A 56, 4452 (1997)

\bibitem{Wootters-1998} William K. Wootters
Phys. Rev. Lett. 80, 2245

\bibitem{Vidal-2002} G. Vidal and R. F. Werner
Phys. Rev. A 65, 032314 (2002)

\bibitem{Liu-2024} Junpeng Liu, Miaomiao Zhao, Yun-Tong Yang, and Hong-Gang Luo
Phys. Rev. A 109, 023721 (2024)

\end{thebibliography}
\bibliographystyle{ieeetr}
\end{comment}

\end{document}